\documentclass[twocolumn]{aastex62}

\newcommand{\myemail}{sarah.r.peacock@nasa.gov}

\def\halpha{\mbox{H$\alpha$}}

\def\lya{\mbox{Ly$\alpha$}}

\def\phx{\texttt{PHOENIX}}

\usepackage[T1]{fontenc}
\usepackage{romanbar}
\usepackage{appendix}
\usepackage{graphicx}
\usepackage{color}
\usepackage{mhchem}
\usepackage{multirow}

\accepted{June 10, 2022}
\submitjournal{ApJ}

\shortauthors{Peacock et al.}

\begin{document}
\title{Accurate Modeling of \lya\ Profiles and their Impact on Photolysis of Terrestrial Planet Atmospheres}

\email{\myemail}

\author[0000-0002-1046-025X]{Sarah Peacock}
\affil{NASA Goddard Space Flight Center, Greenbelt, MD 20771, USA}
\author[0000-0002-7129-3002]{Travis S. Barman}
\affiliation{University of Arizona, Lunar and Planetary Laboratory, 1629 E University Boulevard, Tucson, AZ 85721, USA}
\author[0000-0002-6294-5937]{Adam C. Schneider}
\affil{US Naval Observatory, Flagstaff Station, 10391 West Naval Observatory Road, Flagstaff, AZ 86002-8521, USA}
\affil{3 Department of Physics and Astronomy, George Mason University, 4400 University Drive, MSN 3F3, Fairfax, VA 22030, USA}
\author[0000-0003-1906-5093]{Michaela Leung}
\affiliation{Department of Earth and Planetary Sciences, University of California, Riverside, California, 92521}
\author[0000-0002-2949-2163]{Edward W. Schwieterman}
\affiliation{Department of Earth and Planetary Sciences, University of California, Riverside, California, 92521}
\affiliation{Blue Marble Space Institute of Science, Seattle, WA, USA}
\author[0000-0002-7260-5821]{Evgenya L. Shkolnik}
\affil{School of Earth and Space Exploration, Arizona State University, Tempe, AZ 85281, USA}
\author[0000-0001-5646-6668]{R. O. Parke Loyd}
\affil{Eureka Scientific, 2452 Delmer Street Suite 100, Oakland, CA 94602-3017, USA}

\begin{abstract}

Accurately measuring and modeling the Lyman-$\alpha$ (\lya; $\lambda$1215.67 \AA) emission line from low mass stars is vital for our ability to build predictive high energy stellar spectra, yet interstellar medium (ISM) absorption of this line typically prevents model-measurement comparisons. \lya\ also controls the photodissociation of important molecules, like water and methane, in exoplanet atmospheres such that any photochemical models assessing potential biosignatures or atmospheric abundances require accurate \lya\ host star flux estimates. Recent observations of three early M and K stars (K3, M0, M1) with exceptionally high radial velocities ($>$100 km s$^{-1}$) reveal the intrinsic profiles of these types of stars as most of their \lya\ flux is shifted away from the geocoronal line core and contamination from the ISM. These observations indicate that previous stellar spectra computed with the \texttt{PHOENIX} atmosphere code have underpredicted the core of \lya\ in these types of stars. With these observations, we have been able to better understand the microphysics in the upper atmosphere and improve the predictive capabilities of the \texttt{PHOENIX} atmosphere code. Since these wavelengths drive the photolysis of key molecular species, we also present results analyzing the impact of the resulting changes to the synthetic stellar spectra on observable chemistry in terrestrial planet atmospheres.

\end{abstract}


\NewPageAfterKeywords

\section{Introduction}\label{sec:intro}


The H I Lyman-$\alpha$ (\lya; $\lambda$1215.67 \AA) line comprises $\sim$37\%--75\% of the total UV flux (1150--3100 \AA) from most late-type stars \citep{France2013}, but is difficult to analyze because in $>$99\% of cases it is severely affected by absorption from neutral hydrogen in the interstellar medium (ISM) and if observed from low-earth orbit, contamination from geocoronal airglow. Due to this, model-dependent reconstructions are needed to approximate intrinsic \lya\ fluxes. While significant effort goes into these reconstructions, they depend on assumptions about the shape of the line core and the structure of the ISM. \cite{Youngblood16} estimate that both could independently yield $\sim$30\% inaccuracies, with one of the main sources of uncertainty for \lya\ reconstructions being whether or not self-reversal is assumed in the line profiles. In the case of low resolution observations, \cite{Youngblood2022} found that neglecting self-reversal from \lya\ reconstructions of G and K dwarfs can result in overestimations in flux of up to 100\%, and worse for M dwarfs, up to 180\%.

One of the few stars with non-contaminated \lya\ observations is the Sun. Early modeling efforts recognized that departures from local-thermodynamic equilibrium (LTE) are relevant because the line forms across a region where collisions transition from being important to less important, thereby allowing the line source function to deviate from the Planck function. This non-LTE effect is seen most dramatically in the form of a self-reversal in the line core. 
Recent modeling of late-type stars with the \phx\ \citep{Hauschildt1993, Hauschildt2006, Baron2007} atmosphere code have predicted deep self-reversals in the \lya \ line cores with up to 2$\times$ lower flux than reconstructed profiles of the same stars \citep{Peacock2019,Peacock2019b}.

\begin{figure*}[t!]
    \centering
    \includegraphics[width=1.0\textwidth]{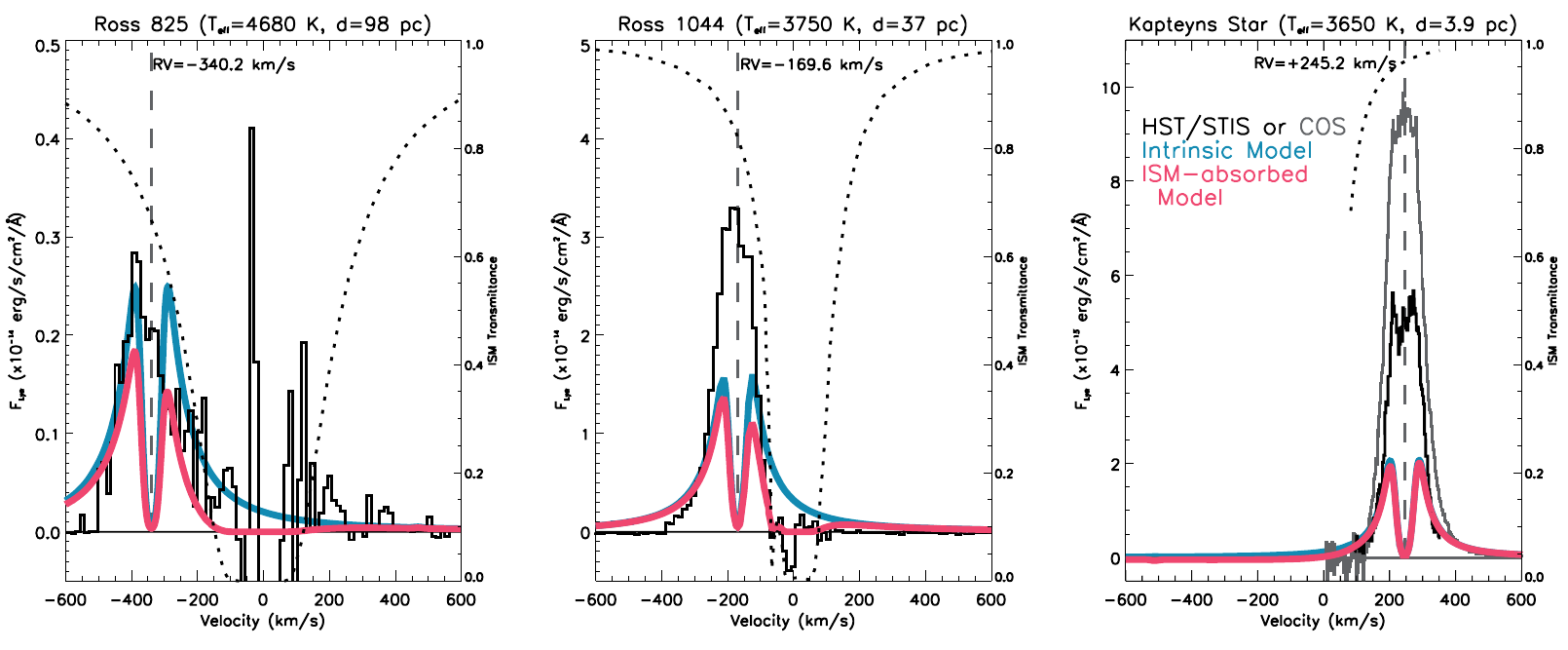}
    \caption{{\it HST} STIS spectra of Ross 825 (K3; \textit{left}), Ross 1044 (M0; \textit{middle}) and Kapteyn's Star (M1; \textit{right}) plotted as thick black lines reproduced from \cite{Schneider2019} and \cite{Youngblood2022}. An altered version of the {\it HST} COS spectrum of Kapteyn's Star from \cite{Guinan2016} is plotted as the grey curve in the third panel. This line profile was produced by mirroring the right-hand side of the stellar component of the \lya \ observation. The large RVs of Ross 1044 (-169.6 km s$^{-1}$), Ross 825 (-340.2 km s$^{-1}$), and Kapteyn's Star (+245.2 km s$^{-1}$) allow their stellar \lya\ emission lines to be well separated from the contaminating geocoronal airglow emission, with little \lya\ flux scattered by the ISM. Model ISM transmittance curves are plotted as black dotted lines \citep{Schneider2019, Youngblood2022}. The blue profiles are the intrinsic PHOENIX model profiles, which when multiplied by the ISM transmittance curves yield the red profiles. These red profiles should reproduce the solid black curves, but severely underestimate the core flux.}
    \label{fig:before}
\end{figure*}

Though it cannot be directly observed for nearly all late-type stars, there does exist one complete \lya\ spectrum, that of Kapteyn's star, an M1 subdwarf with RV=254 km s$^{-1}$ \citep{Guinan2016, Youngblood2022}. There are two additional stars, Ross 1044 (M0) and Ross 825 (K3) that have the majority of their \lya\ flux shifted out of the geocoronal line core and contamination from the ISM because of their exceptionally large RVs ($>$150 km s$^{-1}$) \citep{Schneider2019}. These observations reveal either no self-reversals or very slight ones in the line cores, indicating that previous \phx\ models under predict the \lya \ line core in these stars by a factor of $\sim$2 (Figure \ref{fig:before}).


Accurately measuring and modeling the \lya\ emission lines from low mass stars is vital for our ability to build predictive stellar spectra. The hydrogen spectrum, and in particular \lya, plays an important role in the modeling of low-mass stellar atmospheres for two main reasons: 1) the ionization balance of H I/ H II in the outer layers partly determines the atmospheric structure, and 2) the core of \lya\ is very sensitive to these layers (chromosphere and transition region (TR)) where poorly understood non-radiative heating processes affect the atmospheric structure \citep{Short1997}. The bulk of the extreme ultraviolet (EUV) flux spanning the Lyman continuum ($\leq$912 \AA) is formed in these same outer layers, and therefore, improving the modeling of the \lya\ line directly translates to improvements in modeling the EUV spectrum. Further, accurate stellar \lya\ flux estimates are required by photochemical models assessing exoplanet atmospheric abundances as \lya\ controls the photodissociation of important molecules, including H$_2$O and CH$_4$ \citep[e.g.,][]{Rugheimer2015}. 

For these reasons, \lya \ has been extensively studied over the past several decades. Much of the previous observational work has focused on \lya\ line wings, which are more generally accessible than the complete intrinsic profile since optically thick hydrogen absorption located in the intervening ISM attenuates photons in the \lya \ core. Analyses of observed \lya\ wings have informed us that the line width is directly connected to chromospheric heating, effective temperature, surface gravity, and elemental abundance \citep{Ayres1979, Linsky1980}. \cite{Ayres1979} found that the widths of chromospheric lines are controlled by the stellar temperature distribution more than chromosphere dynamics or magnetic heating. Specific to \lya, \cite{gayley1994} determined that the width of this line is largely controlled by the electron density in the chromosphere, where the broad lines of \lya \ form. Recent work by \cite{Youngblood2022} analyzed \lya\ observations of high RV stars (RV $\sim \pm$ 84$-$245 km s$^{-1}$) and confirmed that this particular chromospheric line width is correlated with surface gravity and effective temperature in addition to the depth of the self-reversal, which decreases with increasing surface gravity.

Early modeling efforts investigated the formation of \lya\ as a function of atmospheric conditions and chromospheric structure. However, the approximations that were typically assumed and the inability to compare to intrinsic \lya\ line profiles left an incomplete understanding of the line formation and sets of constraints on the chromospheric structure.

In a series of papers, the Armagh group \citep{Houdebine1994I, Houdebine1994II, Houdebine1995III, Houdebine1997VI} explored the detailed line formation physics of the hydrogen spectrum in an extensive grid of chromospheric models of early M dwarfs. They found that the temperature break between the chromosphere and transition region (T$_b$) is the main parameter that influences the line width and wings of \lya, but that this temperature break zone is only important for high pressure atmospheres. The densest models produce broad profiles with no self-reversal and strong wings, but as the transition region pressure decreases, these lower pressure models present with extended wings and a self-reversal appears and strengthens. They also found that \lya\ line profiles and fluxes are not sensitive to changes in the temperature minimum, turbulent velocity, nor rotational broadening.

These initial Armagh group studies used the non-LTE radiative transfer code of \cite{Carlsson1986} to reproduce hydrogen emission profiles of M dwarf stars, including high resolution H$\alpha$ and H$\beta$ profiles and the ratio of \lya\ to H$\alpha$ surface fluxes. Their computations used a 16 level hydrogen atom and assumed Complete Frequency Redistribution (CRD). \cite{Short1997} refined these studies using the \phx\ model atmosphere code and included additional physics, specifically adding line blanketing (via the contribution of numerous chromospheric atomic and molecular lines) in the background radiation field. They also treated additional species in non-LTE. In examining the effects of line blanketing on the computation of the hydrogen spectrum, they found that the predicted equivalent width of \lya\ can be reduced by $\sim$0.3 dex and the corresponding line flux increased by a factor of five, concluding that careful treatment of background opacity is important when modeling the hydrogen spectrum. 

\cite{Falchi1998} used the \texttt{Pandora} program \citep{Avrett1984} to further explore the impact of common approximations assumed in chromospheric models: the omission of line blanketing to the opacity, assuming CRD in the \lya\ line, and the assumption that minority species are in LTE. They found that neglecting line blanketing in the background opacities strongly effects the \lya\ line center intensity and also that Partial Redistribution (PRD) effects are very important for computing \lya\ in cool stars. When assuming minority species in LTE versus non-LTE, there was noticeable change in continuum emission shortward of 2600 \AA, but they did not find any effect on the \lya\ line profile. The increased non-LTE species set used for this study included 8 levels of H and 91 levels of 11 ionization stages of He, C, Fe, Si, Ca, Na, Al, and Mg. This set was found to be incomplete by \cite{Fuhrmeister2005}, who concluded that the non-LTE treatment of C, N, and O has a significant influence on the amplitude of hydrogen emission lines in M dwarfs. Recently, \cite{Peacock2019} further increased the non-LTE species set to a total of 62 ionization stages for most light elements up to Ni and added PRD to \phx, confirming that PRD effects are important for reproducing the wings of \lya \ line profiles of old late-type M stars.

These previous studies have made great strides in understanding the formation of the \lya \ line in low mass stars, but none have been able to draw concrete conclusions about the intrinsic profiles of these stars due to the lack of observational/empirical constraints. Now equipped with new observations of high radial velocity stars, over 70\% of the intrinsic \lya\ flux is visible. We use these data to further improve our understanding of the microphysics in the upper atmospheres of low mass stars and, in the process, improve the predictive capabilities of stellar atmosphere models computing EUV spectra. In Section \ref{sec:models} we explain the specifications in our model set up and the stellar observations used as empirical guidance. In Section \ref{sec:analysis} we describe a variety of analyses conducted to test the sensitivity of the \lya\ line core to different components of the model. We present our results in Section \ref{sec:results}. In Section \ref{sec:discussion} we discuss the effects of age and activity on the computation of the \lya\ line and how correctly estimating \lya\ flux in synthetic stellar spectra impacts the modeling of observable chemistry in terrestrial planet atmospheres. We summarize our findings and describe future work in Section \ref{sec:summary}.

\section{Model Set Up}\label{sec:models}

\begin{figure}[t!]
    \centering
    \includegraphics[width=1.0\linewidth]{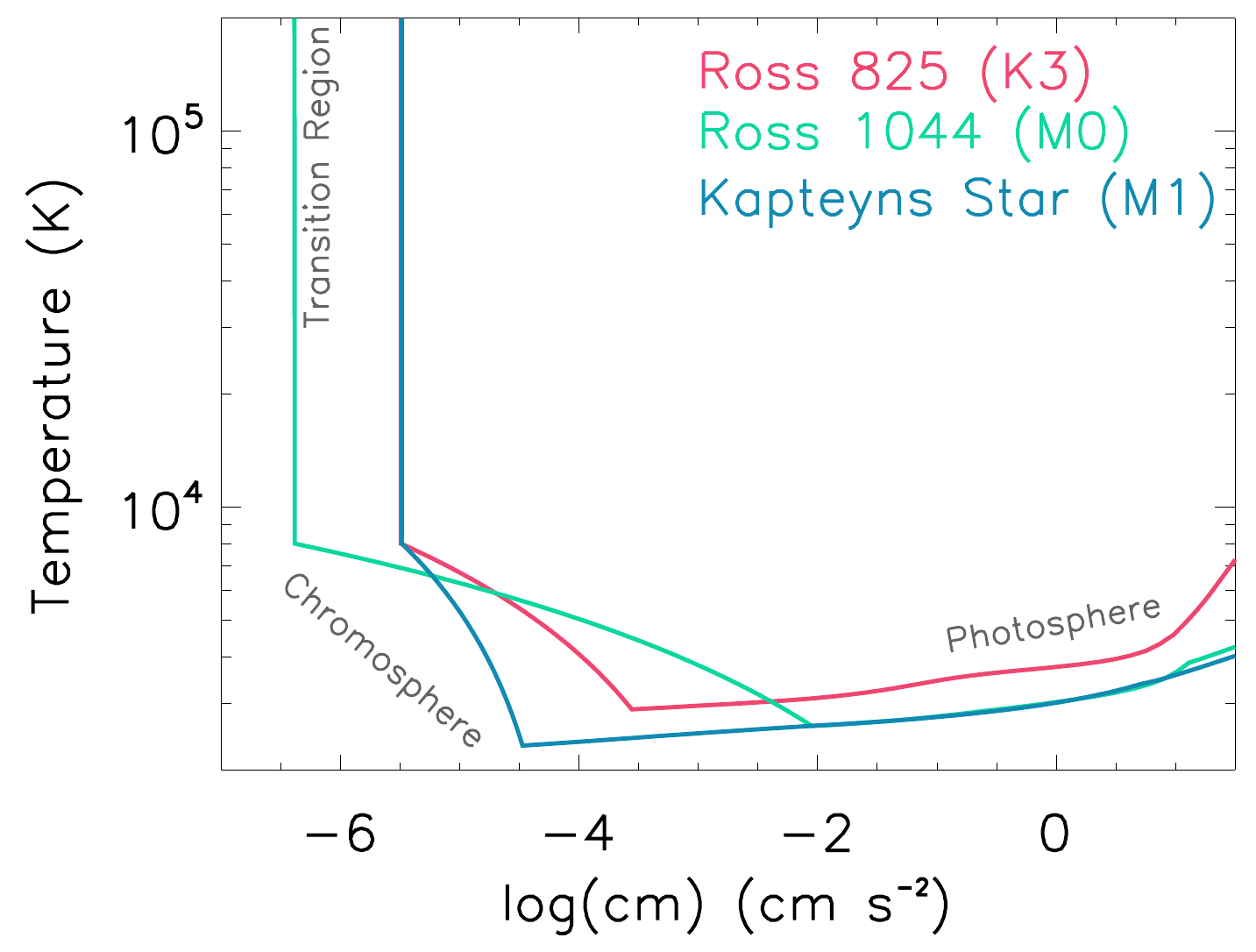}
    \caption{Temperature-column mass structures for models of Ross 825 (red), Ross 1044 (green), and Kapteyn's Star (blue).}
    \label{fig:cmt}
\end{figure}

We computed models with the \phx\ atmosphere code to reproduce the intrinsic \lya\ profiles of Kapteyn's Star, Ross 1044, and Ross 825. For each modeled star, we built photospheric structures defined by literature values of effective temperature ($T_{\rm eff}$), mass (M$_\star$), surface gravity (log($g$)), and metallicity ([Fe/H]) (Table \ref{tab:stelprops}). To the photosphere, we added ad hoc chromospheric and transition region structures that have linear temperature rises with log(column mass) up to 2 $\times$ 10$^5$ K. This maximum temperature is above the range at which \lya\ forms (2 $\times$ 10$^3$ -- 8 $\times$ 10$^4$ K) and is consistent with previous similar modeling efforts \citep{Peacock2019b,Peacock2020}. The temperature at the top of each chromosphere ranges from 7000 -- 8000 K, determined by the point at which hydrogen becomes fully ionized and the atmosphere becomes thermally unstable. The upper atmospheric structures (Figure \ref{fig:cmt}) were adjusted to simultaneously reproduce high resolution \textit{Hubble Space Telescope} (\textit{HST}) \lya\ observations and \textit{Galaxy Evolution Explorer} (GALEX) NUV photometry of each star (Table \ref{tab:fluxes}, Section \ref{subsec:obs}). 

\begin{deluxetable*}{lcccccc}[t!]
    \tablecaption{Stellar Properties \label{tab:stelprops}} 
    \tablehead{
    \colhead{Property} & \colhead{Ross 825} & \colhead{Ref.} & \colhead{Ross 1044} & \colhead{Ref.} & \colhead{Kapteyn's Star}& \colhead{Ref.}
    }
    \startdata
    Spectral Type           & K3&1   & M0&2        & sdM1.0&3\\
    Age (Gyr)               & $>$10&2 & $>$10&2    & 11.5$^{+0.5}_{-1.5}$&4 \\
    RV (km s$^{-1}$)        & -340.17 $\pm$ 0.67 &5  &-169.55 $\pm$ 1.80 &5 & +244.99$\pm$0.18&5\\
    Distance (pc)           & 98.26 $\pm$ 0.18 &5 & 37.44 $\pm$ 0.03 &5 & 3.93$\pm$0.01&5 \\
    $T_{\rm eff}$ (K)       & 4680 $\pm$ 177& 6& 3754 $\pm$ 95& 7 & 3570 $\pm$ 80 (3650) & 8\\
    log($g$) (cm s$^{-2}$)  &4.75& 9  &4.99& 9 &4.96$\pm$0.13& 10 \\
    $R_\star$ ($R_\sun$)    & 0.83 & 11 & 0.53 & 11 &  0.29$\pm$0.025 & 10\\
    $M_\star$ ($M_\sun$)    & 0.55& 9 & 0.3& 9 & 0.28 $\pm$0.01 &10 \\
    $[$Fe/H$]$ & -1.28& 12 & -1.01 $\pm$ 0.21& 13 & -0.86 $\pm$ 0.05&14\\
    \enddata
    \tablecomments{The BHAC97 models \citep{Baraffe1997} were used to obtain the surface gravity and mass for Ross 825 and Ross 1044. We took a rounded average between two rows (corresponding to $T_{\rm eff}$=4377 K and 5068 K) to determine the values for Ross 825. We scaled the models by $R_{\star}^2/d^2$ using literature values for radius for Ross 825 (0.57$^{+0.15}_{0.07}$ $R_\sun$, \citealt{Gaia2018}) and Ross 1044 (0.38$\pm$0.03 $R_\sun$, \citealt{Newton2015}), but this resulted in uniform offsets between measured and synthetic optical and IR photometry for the stars. In order to align the models with the observations, the radii had to be increased to the values listed in this table.}
    \tablerefs{(1) \cite{Bidelman1985};
    (2) \cite{Schneider2019};
    (3) \cite{Hawley1996};
    (4) \cite{Kotoneva2005}/\cite{Wylie2010};
    (5) \cite{Gaia2021};
    (6) \cite{Stassun2018};
    (7) \cite{Newton2015};
    (8) \cite{Anglada2014};
    (9) \cite{Baraffe1997};
    (10) \cite{Segransan2003};
    (11) this work;
    (12) \cite{Schuster2006};
    (13) \cite{Newton2014};
    (14) \cite{Woolf2005}}
\end{deluxetable*}

\begin{deluxetable*}{lcccc}[ht]
    \tablecaption{Observed and Modeled Lyman $\alpha$ and GALEX NUV Fluxes \label{tab:fluxes}}
    \tablehead{
    \colhead{Star} & \multicolumn{2}{c}{Observed} & \multicolumn{2}{c}{Modeled} \\
    & Ly$\alpha$ & NUV& Ly$\alpha$ & NUV
    }
    \startdata
    Ross 825        & 2.65$^{+0.38}_{-0.76} \times$ 10$^{-15}$\tablenotemark{ a}& (1.37 $\pm$ 0.12) $\times$ 10$^{-12}$& 2.79 $\times 10^{-15}$ & 1.45 $\times 10^{-12}$\\
    Ross 1044       & 2.54$^{+0.14}_{-0.11} \times$ 10$^{-14}$\tablenotemark{ a}& (1.47 $\pm$ 0.46) $\times$ 10$^{-13}$& 1.59 $\times 10^{-14}$ & 1.53 $\times 10^{-13}$\\
    \multirow{ 2}{*}{Kapteyn's Star}  & 2.88$^{+0.16}_{-0.08} \times$ 10$^{-13}$\tablenotemark{ b}& (1.09 $\pm$ 0.01) $\times$ 10$^{-12}$& 3.25 $\times$ 10$^{-13}$ & 1.3 $\times$ 10$^{-12}$\\
      & 5.27 $\pm$ 0.26 $\times$ 10$^{-13}$\tablenotemark{ c}& & \\
    \enddata
    \tablecomments{Fluxes are in units of (erg s$^{-1}$ cm$^{-2}$). $F_{\rm Ly\alpha}$ was calculated from 1214.7 -- 1216.7 \AA, $F_{\rm NUV}$ was calculated over the \textit{GALEX} NUV filter bandpass (1687 -- 3010 \AA). All values of $F_{\rm Ly\alpha}$ are the intrinsic line fluxes without ISM absorption.} 
    \tablerefs{(a) \cite{Schneider2019}; (b) \cite{Youngblood2022}; (c) \cite{Guinan2016}}
\end{deluxetable*}

The \lya\ profiles were computed assuming Voigt functions. We also adopted the findings from the aforementioned studies by including line blanketing in the background opacities, computing \lya\ with PRD, and including a robust non-LTE species set. For the line blanketing, we included a total of 15,332 bound-free transitions and 233,871 bound-bound transitions. For the non-LTE calculations, we took into account 15,355 levels for 73 specifically considered ionization stages of the most abundant elements in the Sun\footnote{non-LTE species set in each model: \ion{H}{1}, He {\small\rmfamily I -- II\relax}, C O {\small\rmfamily I -- IV\relax}, N {\small\rmfamily I -- IV\relax}, O {\small\rmfamily I -- IV\relax}, Ne {\small\rmfamily I -- II\relax}, Na {\small\rmfamily I -- III\relax}, Mg {\small\rmfamily I -- IV\relax}, Al {\small\rmfamily I -- III\relax}, Si {\small\rmfamily I -- IV\relax}, P {\small\rmfamily I -- II\relax}, S {\small\rmfamily I -- III\relax}, Cl {\small\rmfamily I -- III\relax}, Ar {\small\rmfamily I -- III\relax}, K {\small\rmfamily I -- III\relax}, Ca {\small\rmfamily I -- III\relax}, Ti {\small\rmfamily I -- IV\relax}, V {\small\rmfamily I -- III\relax}, Cr {\small\rmfamily I -- III\relax}, Mn {\small\rmfamily I -- III\relax}, Fe {\small\rmfamily I -- VI\relax}, Co {\small\rmfamily I -- III\relax}, Ni {\small\rmfamily I -- III\relax}}, including a 30 level hydrogen atom. 

We tested the effects of varying the microturbulent velocity distribution throughout the atmosphere and confirmed the findings of \cite{Falchi1998} that changes in this parameter do not affect the \lya\ line profile. This is because \lya\ is a broad line and other broadening mechanisms are much more important than the Doppler effect. We ultimately used a microturbulent velocity distribution of 2 km s$^{-1}$ in the photosphere and a slope in the chromosphere and transition region that is a fraction of the sound speed (0.35$\times$ $v_{sound}$) in each layer, with a maximum velocity capped at 10 km s$^{-1}$, as originally done in \cite{Fuhrmeister2005} and continued in our previous modeling of low mass stars \citep{Peacock2019,Peacock2019b,Peacock2020}.

With this initial model set up, we were able to reproduce the observed \lya\ line widths but not the line core (Figure \ref{fig:before}). In Section \ref{sec:analysis} we detail various analyses performed to quantify their effects on the intensity of the \lya\ core flux and ultimately match the complete observed line profile.

\subsection{Empirical Guidance}\label{subsec:obs}
To constrain the temperature structure in the chromosphere and transition region, we guided the models to simultaneously match \textit{HST} \lya\ observations and \textit{GALEX} NUV photometry. 

The \textit{HST} observations for Ross 825 and Ross 1044 were taken with the G140M grating on the Space Telescope Imaging Spectrograph (STIS) \citep{Schneider2019}. Kapteyn's Star has been observed twice, both with the STIS/E140M grating \citep{Youngblood2022} and with the lower resolution G130M grating on the Cosmic Origins Spectrograph (COS) \citep{Guinan2016}. There is a large difference in the computed \lya\ fluxes from two these measurements, the lower resolution COS observation yielding a line flux that is 1.85$\times$ that of the higher resolution STIS observation. \cite{Youngblood2022} explain that this difference could be astrophysical in nature or a result of a STIS flux calibration issue. Since the source of the flux inconsistency between measurements is not definitive and the main purpose of this work is to accurately model the line shape of \lya, we choose to match our model to the higher resolution STIS observation. The high RVs of the stars correspond to a \lya\ shift of 1.37 \AA, 0.68 \AA, and 0.99 \AA\ for Ross 825, Ross 1044, and Kapteyn's Star, respectively. These wavelength shifts expose between 63 -- 95 \% of the intrinsic profiles. To compare to these measurements, we convolved our models to the resolution of the observations, accounted for the radial velocity shifts, and multiplied by the ISM transmittance curves shown in Figure \ref{fig:before} (dotted lines).

The \textit{GALEX} NUV photometry for these stars were not taken contemporaneously with the \textit{HST} observations, however, the sample stars are old ($>$10 Gyr), metal deficient ([Fe/H]$<$-0.8), and optically inactive (H$\alpha$ equivalent widths $<$1.0 \AA) \citep{Reid1995,Melbourne2020}, suggesting that the stars were likely emitting similar levels of UV flux during both observations \citep{Houdebine1997VI}. Recent work by \citealt{See2021} found that photometric variability amplitude and metallicity are positively correlated, meaning metal-poor stars, such as those in our sample, are less magnetically active. We therefore are not concerned about the non-contemporaneity of the UV observations used as empirical guidance for our models.


\section{Analysis}\label{sec:analysis}

As described in the previous section, our initial models produced \lya\ profiles that match the wings of the observations well, but underpredicted the core flux by almost a factor of 2. Here we describe a series of analyses conducted to quantify the effect of various model components on the computation of the line profile:

\begin{figure}[t!]
    \centering
    \includegraphics[width=0.93\linewidth]{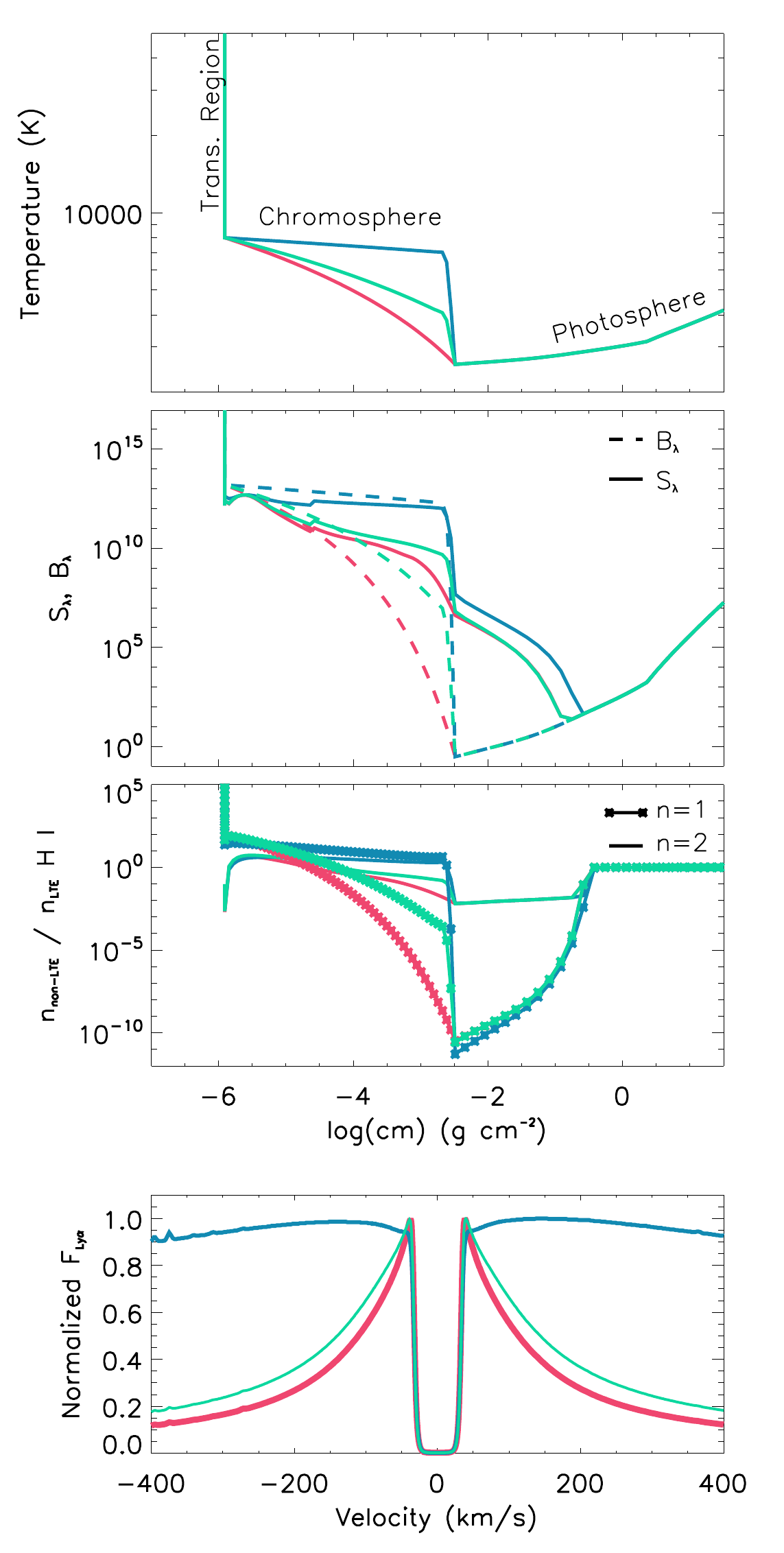}
    \caption{\textit{Top:} Varying temperature structures in the chromospheric layers: our initial set-up with a linear rise (\textit{red}) is modified by adding a steep rise in the lower chromosphere (\textit{green}, \textit{blue}). \textit{Top-middle:} The source (S$_\lambda$, \textit{solid}) and Planck (B$_\lambda$, \textit{dashed}) functions for each model at the center wavelength of \lya\ are plotted in corresponding colors. \textit{Bottom-middle:} The ratios of non-LTE to LTE number density (departure coefficients) for hydrogen for the n=1 (\textit{asterisk}) and n=2 (\textit{solid}) states. Adding nonlinearity to the chromospheric temperature rise results in the source function being more closely tied to the planck function in the mid-chromosphere and larger deviations towards the upper-chromosphere. In all three cases, there is a downturn in both the source function and departure coefficients near the chromosphere-TR boundary. \textit{Bottom:} Normalized \lya\ line profiles. Adding a temperature plateau to the chromosphere increases the \lya\ line width, but does not impact the line core flux.}
    \label{fig:nonlinear}
\end{figure}

\subsection{Effects of Temperature Structure}

First, we explored the impact of the simplifying assumption made in our ad hoc temperature structure, where the chromospheric temperature rise increases linearly with log(column mass). \cite{Houdebine1994I} found that model atmospheres of low mass stars with constant temperature gradients in the chromosphere as a function of log(column mass) best reproduce the observed profiles of hydrogen lines. This finding has been confirmed in many studies including those by \cite{Andretta1997}, \cite{Short1998}, \cite{Fuhrmeister2005}, and \cite{Peacock2019b} where synthetic spectra constructed with linear chromospheric structures simultaneously reproduce the observed UV continuum and many line profiles of cool stars. Other chromosphere models by e.g., \cite{Fontenla2016} and \cite{Tilipman2021} have a steep temperature rise in the lower chromosphere followed by a near-constant temperature plateau in the upper chromosphere similar to the solar structure. The temperature plateau results from the balance of radiative losses with non-radiative heating, and is where singly ionized metals are the dominant stages of ionization \citep{Linsky2017}.

We identified the linearity of the chromospheric temperature structure as a potential source of the discrepancy because the core of \lya\ forms at temperatures near the upper chromosphere and lower transition region \citep{Sim2005} and so, by increasing the temperature in the upper chromosphere, there could be a change in the computed line profile. For this test, we computed models with identical prescriptions in the photosphere and transition region, maintaining the same temperature minimum and temperature at the base of the transition region, but varying profiles in the chromosphere (top panel, Figure \ref{fig:nonlinear}). We used the solar chromospheric structure as guidance, initiating a steep temperature gradient in the lower chromosphere that transitions to a shallower gradient (green) or near-constant temperature plateau (blue). In the second panel of Figure \ref{fig:nonlinear}, we plot radiative quantities for the center wavelength of \lya. In each case, the source function is increasing with temperature in most of the chromospheric layers and the atmosphere is relatively close to thermal equilibrium. These are the layers over which the wings of \lya\ are forming and so the line profiles present with wings in emission (bottom panel). The differences between the models in the lower-to-mid chromosphere results in increasingly inflated wings as the initial chromospheric temperature rise extends to higher temperatures.

The \lya\ core forms near the chromosphere-TR boundary, where departures from LTE are large (third panel) and the emerging photons are no longer coupled to the local temperature. In all three models, the source function in these layers turns over, decreasing with the rising temperature and resulting in deeply self-reversed line cores. With these results, we conclude that the simplifying assumption of a linear rise with log(column mass) in the chromosphere does not impact the core of \lya\ and we maintain that this general temperature prescription produces spectra with the best overall match to both spectroscopic and photometric UV observations.

\begin{figure}[h!]
    \centering
    \includegraphics[width=0.92\linewidth]{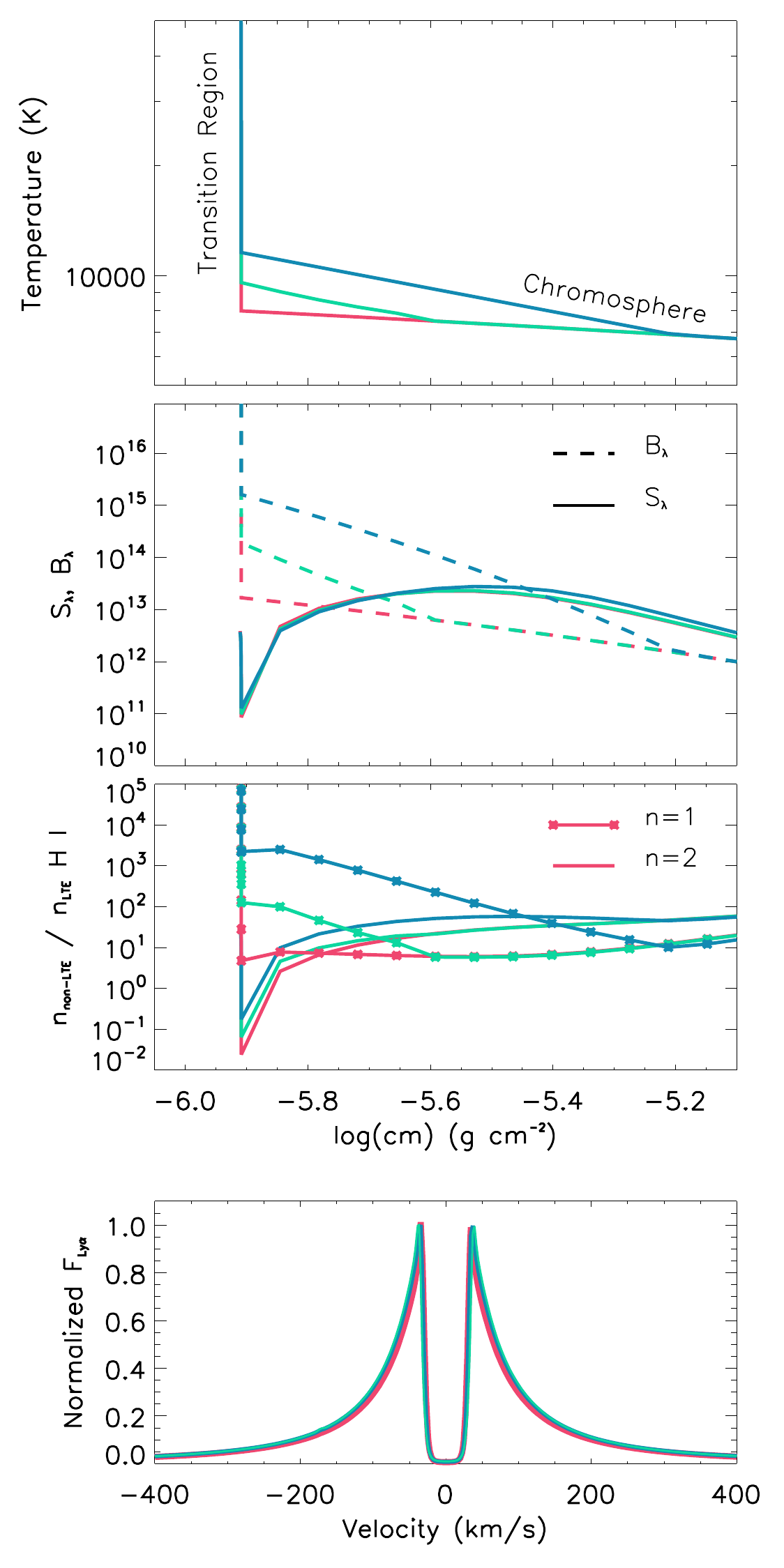}
    \caption{\textit{Top:} A zoomed in view of models with the same general temperature structure with linear temperature rises in the chromosphere and TR, but with increasing degrees of curvature at the chromosphere-TR boundary. Our initial model set-up is shown in red. \textit{Top-middle:} Corresponding: source (S$_\lambda$, \textit{solid}) and Planck (B$_\lambda$, \textit{dashed}) functions at the central wavelength of \lya\ and \textit{Bottom-middle:} departure coefficients for the n=1 (\textit{asterisk}) and n=2 (\textit{solid}) states of hydrogen. Adding increasing curvature to the temperature structure at the boundary region reduces the severity of change between the temperature gradients in the chromosphere and transition region. This change results in an increase the n=1 departure coefficients over the boundary layers, but does not significantly impact the source function nor the n=2 departure coefficients. \textit{Bottom:} Adding curvature to the boundary between the chromosphere and transition region does not affect the computed \lya\ profiles.}
    \label{fig:smoothTR}
\end{figure}

\begin{figure*}[t!]
    \centering
    \includegraphics[width=1.0\textwidth]{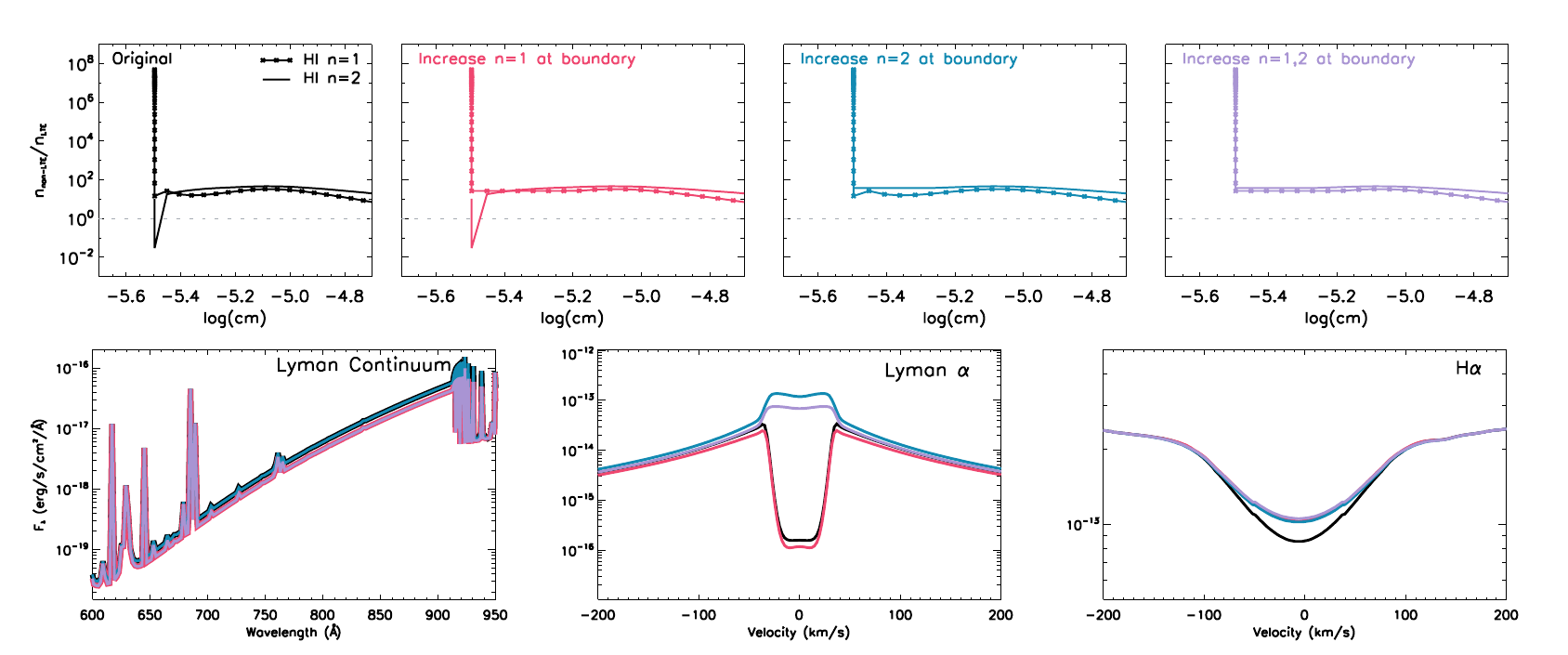}
    \caption{Adjusting the ratios of the non-LTE to LTE number density (n$_{non-LTE}$/n$_{LTE}$) for hydrogen in the n=1 and n=2 states in the layers surrounding the chromosphere-TR boundary affect the flux in the Lyman continuum (n=1$\rightarrow \infty$), \lya\ (n=2$\rightarrow$1), and \halpha{} (n=3$\rightarrow$2). \textit{Top Row}: n$_{non-LTE}$/n$_{LTE}$ for \ion{H}{1}(n=1) (\textit{asterisks}) and \ion{H}{1}(n=2) (\textit{solid}) in the transition region (log(cm)$<$-5.5) and upper chromosphere (log(cm)$>$-5.5). These departure coefficients for the initial \phx\ model set up are shown in the top left panel (\textit{black}). Subsequent panels in the top row have a minimum of n$_{non-LTE}$/n$_{LTE}$ = 3 set for n=1 (\textit{red}), n=2 (\textit{blue}), and both n=1\&2 (\textit{purple}). \textit{Bottom Row:} Output \phx\ spectra corresponding with each departure coefficient adjustment are plotted in matching colors. \textit{Bottom Left:} The Lyman continuum decreases as \ion{H}{1}(n=1) is increased at the chromosphere-TR boundary (\textit{red}, \textit{purple}). \textit{Bottom Middle:} The core of Lyman alpha changes in all three scenarios, with the reversal deepening as \ion{H}{1}(n=1) is increased and filling in as \ion{H}{1}(n=2) is increased. \textit{Bottom Right:} \halpha{} is sensitive to changes in both n=1 and 2, increasing flux in the line core as the values are increased at the boundary.}
    \label{fig:flatten}
\end{figure*}

Next, we examined the temperature structure over the narrow range where the chromosphere and transition region are joined. We investigated this narrow range because it is the specific region where the core of \lya\ forms and where the previous models have a sharp downwards turn in the source function. We considered that there may be a numerical issue resulting from the dramatic change in temperature gradient between the chromosphere ($\nabla$T$_{\rm ch}$ $\simeq$ 10$^6$ K g$^{-1}$ cm$^{2}$) and the transition region ($\nabla$T$_{\rm TR}$ = 10$^8$ g$^{-1}$ cm$^{2}$). 

For this test, we took a model with our initial set-up (linear temperature rises in the chromosphere and TR; top panel, Figure \ref{fig:smoothTR}, red) and smoothed the layers around the chromosphere-TR boundary, adding curvature over increasingly broader ranges of temperature and column mass (Figure \ref{fig:smoothTR}, green and blue). The alterations to the temperature structure over these boundary layers do not impact the source function (top-middle panel, Figure \ref{fig:smoothTR}, solid curves). In each model, the source function traces the Planck function through the chromosphere, but deviates in the upper chromosphere and TR, decreasing with increasing temperature and resulting in the self-reversed core. The increased curvature results in larger departures from non-LTE in the ground state of hydrogen (bottom-middle panel, Figure \ref{fig:smoothTR}), but minimal change to the n=2 state. Ultimately, adding this curvature to the chromosphere-TR boundary does not affect the computed \lya\ line profile. 

In order to see any change in \lya, we had to greatly extend the curvature, smoothing the layers between 6,000 and 20,000 K. In doing so, the entire UV spectrum short-ward of 2500 \AA\ increased in flux by $\sim$2 orders of magnitude, but the \lya\ core flux only showed modest changes in the depth of the central reversal, increasing the line flux by 10\%.\\

\begin{figure*}[t!]
    \centering
    \includegraphics[width=1.0\textwidth]{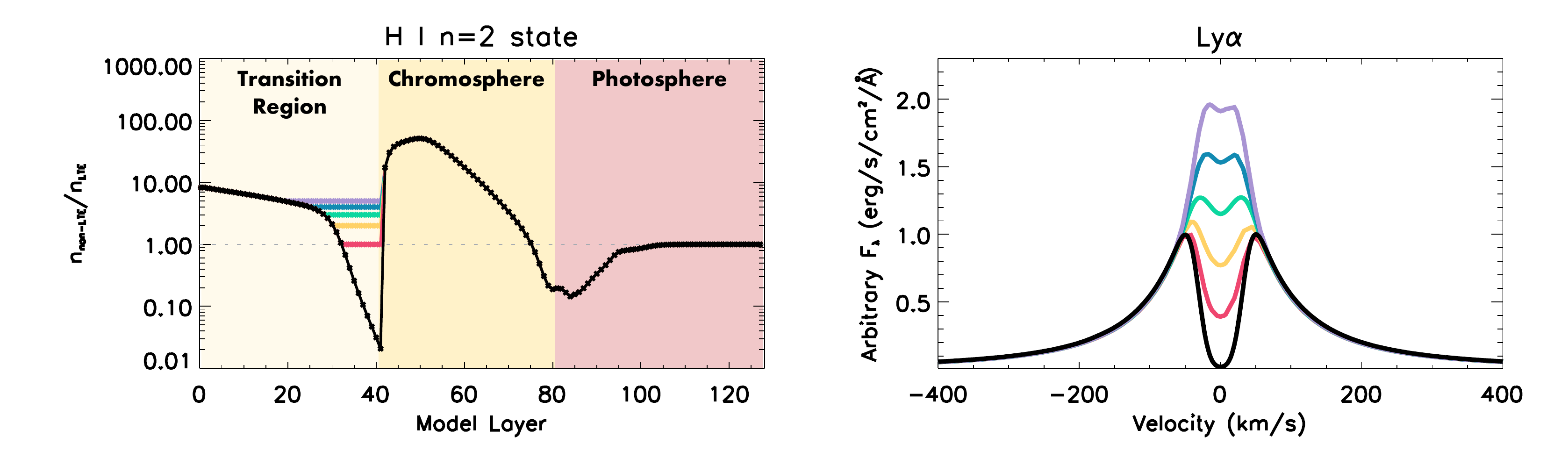}
    \caption{The depth of the self-reversal in the \lya\ line is directly related to n$_{non-LTE}$/n$_{LTE}$ for hydrogen in the n=2 state in the layers surrounding the chromosphere-TR boundary. {\it Left:} Varying minimums for n$_{non-LTE}$/n$_{LTE}$ are set in these layers, shown with different colors. {\it Right:} Corresponding \lya\ line profiles show the sensitivity of the self-reversal to these changes.}
    \label{fig:bicut}
\end{figure*}

\subsection{Departure Coefficients}

The departure coefficients, or the ratios of non-LTE to LTE number density, are a proxy for the population of each level and are required for the calculation of the emissivity and absorption coefficients. The \lya\ line is formed by electron transitions in the hydrogen atom from the level 2 state to ground (n=1). The Lyman continuum, which largely shapes the EUV spectrum shortwards of 912 \AA, results from bound-free transitions from the ground state, while the Balmer series are transitions to the n=2 state, including the H$\alpha$ line (n=3 $\rightarrow$ 2, 6562.5 \AA) and the Balmer continuum (n=2 $\rightarrow \infty$, 3647 -- 6563 \AA).

A commonality observed in all of the models in the previously described temperature structure analyses was that the level 2 state of hydrogen is underpopulated at the base of the transition region. \cite{Falchi1998} also found that this second level of hydrogen was underpopulated for M dwarf chromosphere models computed with the \texttt{Pandora} program \citep{Avrett1984}, but for CRD models of inactive stars, only. Our stars are similarly inactive, however the underpopulation is still apparent when including PRD for the computation of \lya.

In our original model set-up, the departure coefficients for the n=1 state of hydrogen slightly decrease at the chromopshere-TR boundary, while the n=2 state drops by several orders of magnitude (Figure \ref{fig:flatten}, black). We have found that this underpopulation is a direct result of an upward jump in the radiative rates at the onset of the transition region. When manually increasing the ratio for n=1 over all layers in the upper chromosphere where previous downturns occurred (Figure \ref{fig:flatten}, red), the flux in the Lyman continuum decreases and the self-reversal in the \lya\ core deepens. When the same adjustment is made to the n=2 state (Figure \ref{fig:flatten}, blue), the Lyman continuum is unchanged and the flux in the core of \lya\ increases such that the line is nearly in full emission, displaying only a slight self-reversal. Simultaneously increasing the ratios for both n=1 and 2 yields additive results from adjusting n=1 or n=2 alone (Figure \ref{fig:flatten}, purple): the same decreased flux in the Lyman continuum from adjusting just the n=1 state, and a slightly weaker \lya\ profile than that resulting from adjusting just n=2. 

These findings follow as the bound-bound source function can be expressed with departure coefficients as:

\begin{equation}
    S^l_{\nu} \approx \frac{b_u}{b_l}B_{\nu} =\frac{2h\nu^3}{c^2}\frac{\psi/\phi}{\frac{b_l}{b_u}e^{h\nu/kT}-\frac{\chi}{\psi}}
\end{equation}

\noindent where $b_u$ and $b_l$ are the departure coefficients, $n_{non-LTE}/n_{LTE}$, for the upper and lower level, respectively, as defined in \citep{Rutten}. $B_{\nu}$ is the Boltzmann distribution, given in full form on the right hand side of the equation, where $\psi$, $\phi$, and $\chi$ are the profile functions for extinction, emission, and induced emission. Increasing the departure coefficients in the upper level, n=2, increases the source function and results in emission. Increasing the departure coefficients in the lower level, n=1, has the inverse effect.

H$\alpha$ is sensitive to changes to both the n=1 and 2 departure coefficients, increasing flux in the line core as the ratios are increased in the upper chromosphere. This emission line does not directly correspond to electron transitions with the ground state of hydrogen, but adjusting the population in the n=1 state effects the availability of electrons for other transitions, which is why there is still a slight change in flux. In these stars, the Balmer continuum is very weak and these adjustments to the departure coefficients change the total flux in those wavelengths by $<$1\%.

In Figure \ref{fig:bicut}, we analyze the sensitivity of the \lya\ line to the ratio of n$_{non-LTE}$/n$_{LTE}$ for hydrogen in the n=2 state. Our original model is plotted in black with asterisks denoting each model layer. At the chromosphere-TR boundary, there is a sharp decrease of three orders of magnitude in n$_{non-LTE}$/n$_{LTE}$. We have found that the strength of the self-reversal in the core of \lya\ is highly sensitive to the minimum set in the layers surrounding the chromosphere-TR boundary. By setting the minimum to 1.0, the sharp decrease between Model Layer 42 in the chromosphere and Model Layer 41 in the transition region reduces to approximately one order of magnitude and the central reversal in the line profile becomes less severe than in the original model (red and black curves). The difference in \lya\ line flux between these two models is $\sim$10\%. Increasing the minimum to 5.0 (purple curve), the difference in n$_{non-LTE}$/n$_{LTE}$ in these layers decreases to a factor of 3 and the \lya\ line profile is nearly in full emission, with 1.5$\times$ the line flux in the original model.

\begin{figure*}[t!]
    \centering
    \includegraphics[width=1.0\textwidth]{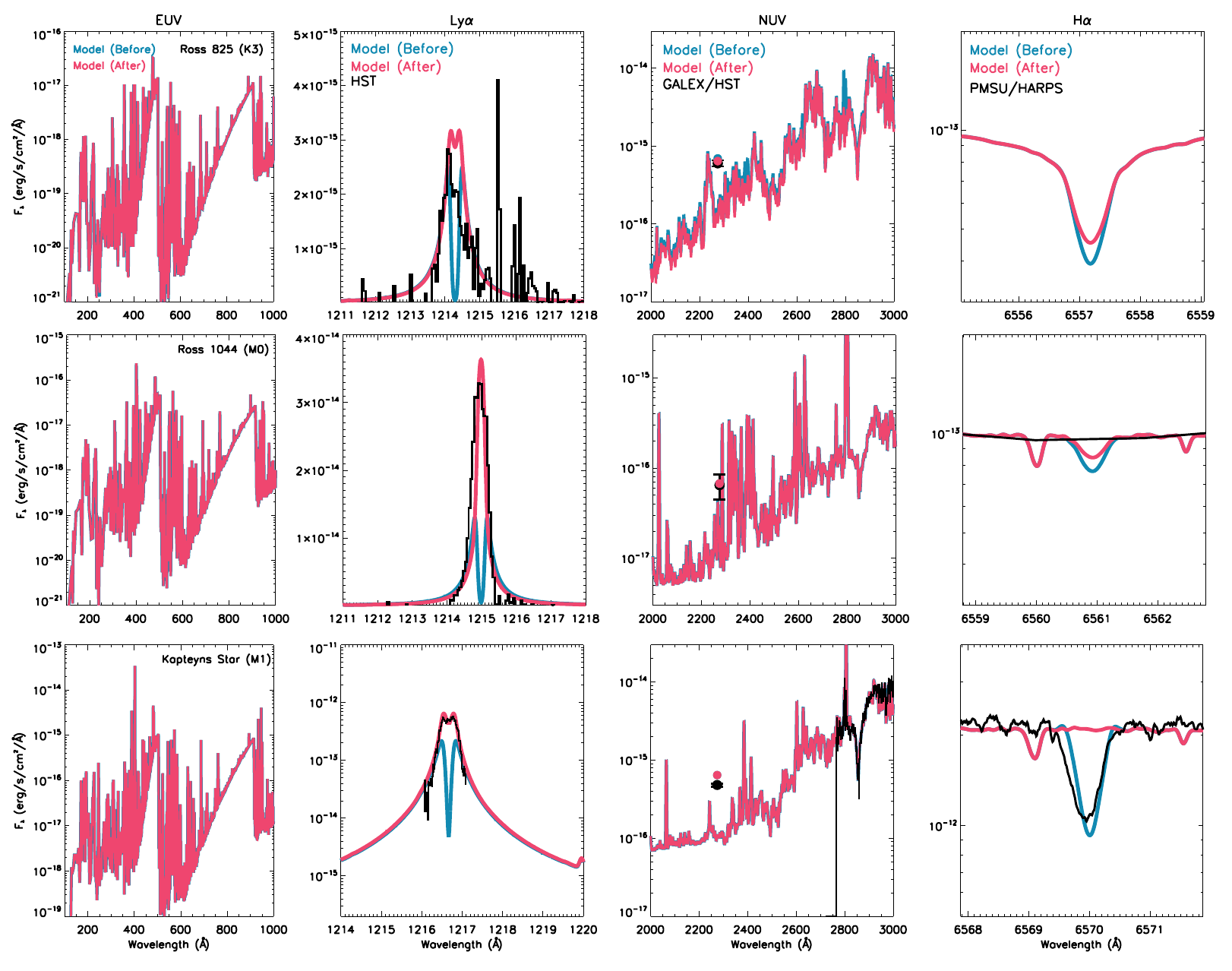}
    \caption{The corrected \phx\ spectra for Ross 825 (top), Ross 1044 (middle), and Kapteyn's Star (bottom) are plotted in red and compared to both the initial \phx\ models (blue) and the \textit{HST}, \textit{GALEX} and \halpha{} measurements of each star (black). The \halpha{} observation for Ross 1044 comes from the Palomar/Michigan State University (PMSU) survey \citep{Reid1995} and yields a flat profile. The Kapteyn's Star \halpha{} observation comes from a HARPS observation taken within 6 months of the \lya\ profile \citep{Anglada-Escude2016b}. There are no available \halpha{} observations for Ross 825. The final model spectra have good agreement with the measured \lya\ profiles, including core flux, and the NUV photometry. When multiplying the final model spectra by the ISM transmittance curves from Figure \ref{fig:before}, the \lya\ profiles overlay the \textit{HST} observations. The comparison with initial \phx\ models show that the adjustments made to the \ion{H}{1}(n=2) departure coefficients result in negligible changes to EUV or NUV wavelengths, only increasing flux in the cores of \lya\ and \halpha{}.}
    \label{fig:beforeandafter}
\end{figure*}

\begin{figure*}[t!]
    \centering
    \includegraphics[width=1.0\textwidth]{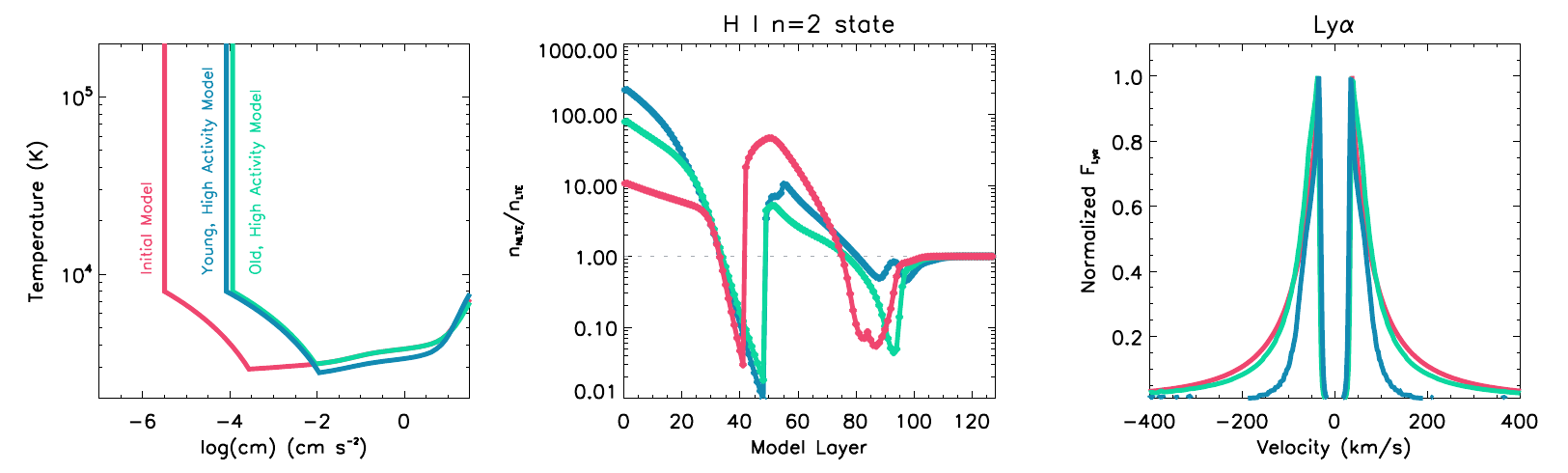}
    \caption{Parameterizations for stellar inactivity (including metallicity and the onset temperature for the chromospheric temperature rise) do not affect the severity of the  decrease in n$_{non-LTE}$/n$_{LTE}$ at the chromosphere-TR boundary for hydrogen in the n=2 state (\textit{middle panel}). We compare our initial model ($>$10 Gyr, [Fe/H]=-1.28, \textit{red}) to a young, high activity model (10 Myr, [Fe/H]=0.0, \textit{blue}) and old, high activity model ($>$10 Gyr, [Fe/H]=-1.28, \textit{green}). The old, high activity model has the same photospheric parameters as the initial model, but with it's chromosphere initiated at higher column mass, similar to that of the young model. \textit{Right:} Normalized Normalized \lya\ line profiles. The corresponding \lya\ profiles present with similarly deep central reversals, but increase in line flux with increasing chromospheric activity and metallicity.}
    \label{fig:highactivity}
\end{figure*}

\section{Results}\label{sec:results}

As the computed \lya\ profiles are highly sensitive to the changes in the departure coefficients of \ion{H}{1}, we were able to reproduce the observations by specifying a tailored minimum for the n=2 state in the layers surrounding the chromosphere-TR boundary for each star. The final computed \lya\ and NUV fluxes for each star are listed in Table \ref{tab:fluxes}, reproducing the observed quantities. To match these observations, the minimum is set to 3 for Ross 825 and Kapteyn's Star, and set to 10 for Ross 1044. We compare the models before and after these minima are set in Figure \ref{fig:beforeandafter}. The difference in \lya\ flux for each star ranges from 35 -- 150\%. The difference in H$\alpha$ flux ranges from 5 -- 20\%. There is no discernible change to any other part of the spectrum besides these two emission lines. This result contradicts a finding in \cite{Houdebine1994I} that Lyman and Balmer series lines can be modelled almost separately, however, it should be noted that by modifying the departure coefficients manually, we are manipulating the occupation numbers and no longer conserving the total number density of hydrogen. This means that the model is no longer physically consistent, however, 99.99\% of the total number density of hydrogen is in the n=1 state (throughout the entire atmosphere and specifically across the chromosphere-TR boundary), so increasing the occupation numbers in the n=2 state does not significantly affect the total number density of hydrogen.

We have found that the cause for the underpopulation of hydrogen in the n=2 state is related to an upward jump in the radiative rates, indicating an issue with the mean intensity caused by missing or incorrect opacities in the code. While we are using a large non-LTE species set, there are still some residual transitions computed in LTE. Removing these lines from the model calculation does not affect the radiative rates, therefore, this is not the cause and we suspect there may be some other missing UV opacity source.

Earlier modelling efforts utilized a \lya\ to H$\alpha$ flux ratio in low mass stars as a diagnostic for determining the thickness of the transition region. \cite{Doyle1990} found that the ratio of observed \lya\ to H$\alpha$ fluxes for M dwarfs is close to 1, using low resolution International Ultraviolet Explorer (IUE) satellite data to extract the \lya\ profiles. While they separated the geocoronal from stellar contributions, they did not correct for interstellar absorption. As a result, these early modeling efforts \citep[e.g.,][]{Houdebine1994I, Short1997} struggled to reproduce the flux ratio, consistently producing models with \lya\ to H$\alpha$ flux ratios between 2 -- 5. Now equipped with \lya\ observations that reveal the intrinsic line flux to guide our models, we find that the continuum normalized \lya \ to H$\alpha$ flux ratio is between 5 -- 20. This discrepancy is likely due to a combination of the initial modelling efforts not accounting for the ISM absorption in addition to our sample stars being very old and optically inactive, resulting in flat \halpha{} profiles.\\

\section{Discussion}\label{sec:discussion}

\subsection{Effects of Stellar Inactivity}

All three of our sample stars are old and inactive, with subsolar metallicity. The low metallicity of these stars results in low electron density at chromospheric temperatures resulting in weak \lya\ emission compared to more solar-like stars, however, the effects of metallicity have not been extensively studied in chromospheric modelling previously. Because these stars are not representative of the general stellar population, we consider the effects of both activity and [Fe/H] on the \lya\ line and the population of hydrogen in the n=2 state.

As FGKM stars age, they increase in effective temperature, but decrease in radius and gravity. They spin more slowly, have reduced dynamo production of magnetic fields, and become less UV active as their chromospheres shift outward to lower pressures \citep{Peacock2020}. Older stars also have lower metallicity than their younger counterparts because they were born in an environment where less metals were available. In Figure \ref{fig:highactivity}, we compare our initial >10 Gyr model from Section \ref{sec:analysis} (red) to a star with same photosphere, but more active upper atmosphere (green) and a young star, defined by lower $T_{\rm eff}$, higher gravity and solar metallicity, plus a more active upper atmosphere (blue). This young star is representative of the initial model star at 10 Myr, obtaining the photospheric parameters from BHAC15 models ($T_{\rm eff}$=4142 K, log(\textit{g})=4.184, [Fe/H]=0.0) \citep{Baraffe2015} and the upper atmospheric structure from a 10 Myr early M star \citep{Peacock2020}.

In both the >10 Gyr model with increased chromospheric activity and the 10 Myr model with solar metallicity, the n=2 state of hydrogen is similarly underpopulated at the the boundary between the chromosphere and transition region. The sharp decrease between the ratios of non-LTE to LTE number density at these layers is comparable to that of the initial model regardless of the choice of chromospheric parameterization or [Fe/H]. As a result, the \lya\ profiles for all three cases present with a deep self-reversals in the line core. The total line flux does increase with activity, increasing by 25\% between the initial model and the old, high activity model, and by a factor of $\sim$10 between the initial and 10 Myr models.

While the underpopulation of hydrogen in the n=2 state is not significantly affected by including the effects of metallicity in the modelling, it is important to include [Fe/H] in these calculations. Keeping all parameters equal except for [Fe/H], there are noticeable changes throughout the computed spectrum. Considering just \lya, an increase in [Fe/H] by 0.3 increases the \lya\ flux by $\sim$25\%, similar to the changes seen by shifting the initial chromospheric temperature rise inwards towards higher column mass.

\subsection{Photochemical Impact on Terrestrial Planetary Atmospheres}

\begin{figure}[t!]
    \centering
    \includegraphics[width=0.45\textwidth]{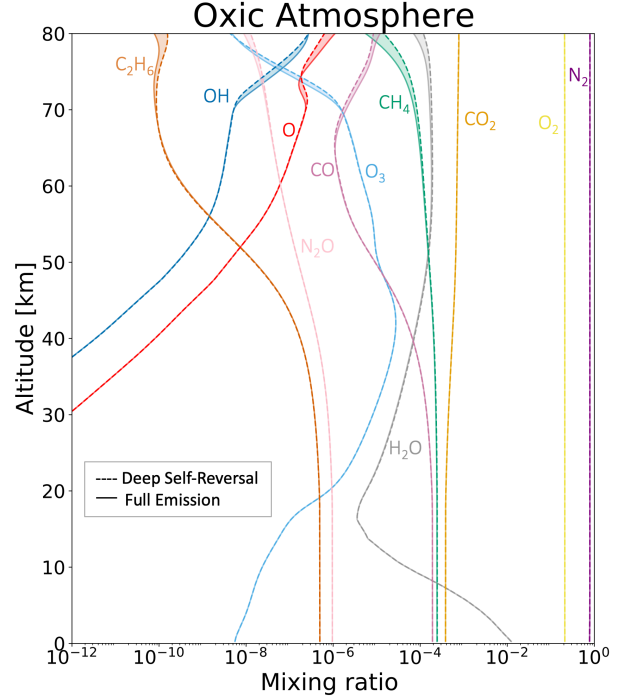}
    \caption{Comparison of gas mixing ratio profiles in an oxic atmosphere for the endmember stellar spectra explored here (full emission and deep self-reversal \lya\ fill cases). At the top of the atmosphere, some species including CH$_{4}$, H$_{2}$O, and N$_{2}$O show slightly diminished concentrations in the full emission case as a result of increased photolysis. Photolysis products CO, O, and OH, show slightly increased concentrations for the same reason. Below $\sim$ 55 km, the profiles are essentially identical. }
    \label{fig:oxic_atmosphere}
\end{figure}

\begin{figure}[t!]
    \centering
    \includegraphics[width=0.45\textwidth]{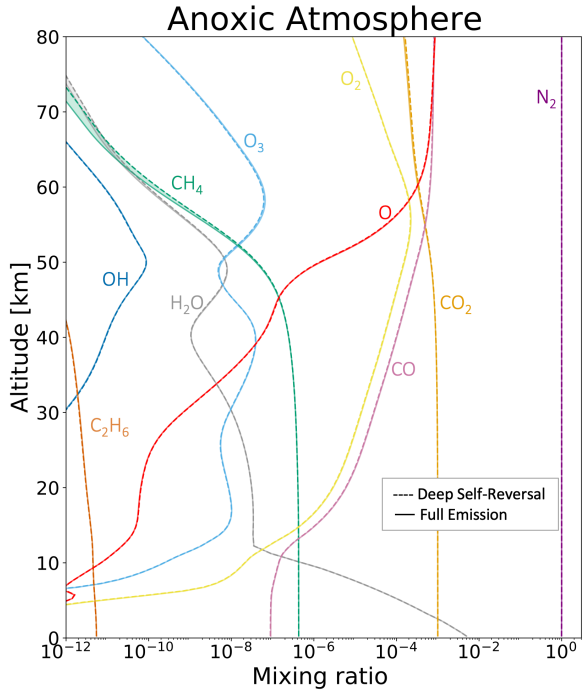}
    \caption{Anoxic atmosphere comparison of full emission and deep self-reversal \lya\ fill cases. Small differences in the \ce{CH4} and \ce{H2O} profiles can be seen in the mesosphere, but the profiles are equal below 50 km. Similar to the oxic scenario shown in Figure \ref{fig:oxic_atmosphere}, the concentrations of CH$_{4}$ and H$_{2}$O are slightly lower at the top of the atmosphere in the full emission case resulting from higher photolyzing flux from \lya fill, while photolysis products (CO, O) are slightly greater in concentration.}
    \label{fig:anoxic_atmosphere}
\end{figure}

Recent work by \cite{Teal2022} quantified the effects of using different quality stellar UV inputs when modeling the photochemical response of exoplanet atmospheres. They found that comparable results are obtained when using either scaling relations or reconstructed proxy spectra as UV input for modeling modern Earth-like planets, but that observable differences occur in the modeled transmission spectra of hazy planets depending on the choice of UV input. They concluded that the biggest effect comes from the UV continuum. Here we take a closer look specifically at the impact of \lya\ on the photochemical response of terrestrial planet atmospheres, and quantify the relative importance of modeling this line correctly.

For this analysis, we used the \texttt{atmos} photochemical model, previously described in \cite{Arney2016}, \cite{Lincowski2018ApJ}, and \cite{Teal2022}, to examine the impact of different fills in the core of \lya\ (similar to those shown in Figure \ref{fig:bicut}). \texttt{Atmos} is a 1-D photochemical model that has been used for a variety of terrestrial exoplanet applications, including to simulate planetary atmospheres similar to those of the modern and early Earth \citep{Arney2016, Teal2022}. We used the most recent publicly available version of \texttt{atmos}\footnote{https://github.com/VirtualPlanetaryLaboratory/atmos}, including all of the modifications described in \cite{Teal2022}. These modifications include a high-resolution internal grid that prevents \lya\ from being spread over too wide a wavelength range. This is particularly important for our analysis in which we are fine tuning the \lya\ profile. 

We explored two photochemical reaction templates: one that replicates the modern O$_{2}$-rich Earth around the Sun (21\% O$_{2}$), and another for an anoxic low methane atmosphere. The "anoxic" atmosphere is not entirely without oxygen; however, its O$_{2}$ content is determined self-consistently by photochemical production and destruction and loss to deposition at the surface, rather than a large fixed mixing ratio. The oxic atmosphere contains 239 reactions and 50 gaseous species while the anoxic case has 392 reactions and 74 gaseous species. The oxic atmosphere case assumes a 288 K surface with the empirical temperature-pressure profile of the modern Earth and water vapor consistent with a surface relative humidity of 80$\%$. Fixed surface flux boundary conditions are assumed for CH$_{4}$ and N$_{2}$O consistent with the biological productivity of the modern Earth. The low-methane, anoxic atmosphere assumes a 275 K surface with a 180 K stratosphere and the same surface relative humidity of of 80$\%$. The fixed surface flux of CH$_{4}$ is based on a reasonable maximum abiotic flux from volcanism and water-rock reactions \citep{Guzman2013}. The oxic and anoxic cases also vary in their assumed CO$_{2}$ concentrations, 400 ppm and 1,000 ppm, respectively. In both cases the surface pressure is 1 bar and N$_{2}$ is used as a filler gas. 

Figure \ref{fig:oxic_atmosphere} shows the differences in an oxic atmosphere exposed to a stellar spectrum with a deeply self-reversed \lya\ profile (Figure \ref{fig:bicut}, black) and the same spectrum with \lya\ in full emission (Figure \ref{fig:bicut}, purple). There are minor changes in the overall mixing ratio profiles, specifically located at the top of the atmosphere. In general, the full emission scenario shows slightly lower concentrations for most gases in the mesosphere (>50km) due to more robust photolysis (with the exception of photolysis products such as CO and O, which are enhanced for the same reason). Gas concentrations in the stratosphere (14-50 km) and troposphere (surface to 14 km) are essentially unchanged. For all atmospheric constituents, the change is less than a factor of two, an amount that does not significantly impact the detectability of spectral features. This is because the locations in the atmosphere with the greatest divergence are of the lowest densities and would contribute minimal opacity in transit transmission observations. 

The anoxic case, shown in Figure \ref{fig:anoxic_atmosphere}, displays similar trends when comparing the deep self-reversal and full emission \lya\ line scenarios. The larger CO$_{2}$ concentration for this case results in more FUV (and specifically \lya) shielding at the top of the atmosphere, muting the photolysis impact for most altitudes. The most notable difference between the deep self-reversal and full emission scenarios is in the mesospheric CH$_{4}$ concentrations. This is due to more robust photolysis at the top of the atmosphere for the full emission scenario, where there is a higher total amount of stellar UV flux. Similar to the oxic case, these changes are less than a factor of two and would not impact the detectability of spectral features in the planetary spectrum due to the low densities at the most impacted altitudes.

When stellar \lya\ is in full emission, the atmospheric profiles for both the anoxic and oxic planets show very slight amounts of increased photolysis at altitudes near the top-of-atmosphere. Overall the atmospheric mixing ratios and potentially detectable spectral features are not significantly impacted by the differing \lya\ profiles explored here. These findings are consistent with those of \cite{Teal2022} who find that their results are most sensitivity to the UV continuum, which is is fixed between scenarios presented here.

\section{Summary/Conclusions}\label{sec:summary}

The \lya\ observations of Ross 825, Ross 1044, and Kapteyn's Star have allowed for mostly unobstructed access to the line cores of \lya, revealing the intrinsic profiles of old, inactive, low-mass stars. Our modeled profiles show more severe self-reversals with earlier spectral type, consistent with previous findings that the \lya\ core reversal depth correlates with surface gravity \citep[e.g.][]{Youngblood2022}. 

Modeling the \lya\ lines of these stars has demonstrated the sensitivity of the core flux to the departures from LTE in the n=2 state of \ion{H}{1} at the boundary between the chromosphere and transition region. These departure coefficients have revealed that there are missing or incorrect opacities in the code and require different minimums to be set in these layers in order to match each \lya\ profile. When the models are adjusted to match the observations, there is an increase in \lya\ flux of 35--150\% and minimal effect on the Balmer series or EUV spectrum. The H$\alpha$ flux decreases by $\leq 20\%$, while the Balmer continuum and EUV spectrum both decrease by $<1\%$.

We analyzed the impact of these stellar spectral differences when modeling the photochemical response of high molecular weight terrestrial exoplanet atmospheres. We found that corresponding changes to spectrally active gases are very small compared to other uncertainties and will not impact the modeling of planetary spectra. 


We also confirmed the using a simplifying assumption of a linear temperature rise with log(column mass) when modeling the stellar chromosphere produces spectra with the best overall match to UV observations. We found that smoothing the temperature structure at boundary of the chromosphere and transition region yielded negligible changes to the computed UV spectrum.

\subsection{Future Work}

The results from this work re-emphasize that the depth of the self-reversal in the core of \lya\ increases with spectral type, however, it is unclear how exactly the line profile changes with uniform steps in $T_{\rm eff}$ of less than 1000 K. This is important to know for improving the accuracy of modeling stellar atmospheres for a broad range of stars. In order to match the observed profiles of the three target stars in this work, tailored adjustments had to be made. This limited sample size did not reveal a clear trend for what the minimum value for n$_{non-LTE}$/n$_{LTE}$ needs to be based on the $T_{\rm eff}$ or gravity of the star, although there may be a correlation with the location of the TR. Increasing this sample size would potentially reveal a correlation between the n=2 departure coefficients and $T_{\rm eff}$ or other stellar parameters that can be used to apply an accurate correction to all models

With the upcoming HST program 16646, we are tripling the observational sample of low-mass stars for which \lya\ can be measured directly (RV $>$ 100 km s$^{-1}$), uniformly sampling stars with $T_{\rm eff}$ from 3400--5500 K in steps of $\sim$500 K. This will enable us to better understand how the \lya\ profile changes with spectral type. \cite{Ayres1979}, \cite{Linsky1980}, and \cite{Youngblood2022} have each connected chromospheric emission line widths as a function of chromospheric heating, $T_{\rm eff}$, surface gravity, and elemental abundance. With these new observations we will determine if the total \lya\ line strength is connected as well.

Reconstructed \lya\ fluxes have been used extensively to produce correlations with other spectral emission lines \citep{Linsky2013, Youngblood2017, Melbourne2020}, broadband UV photometry \citep{Shkolnik2014}, and X-ray fluxes \citep{Linsky2020}. Such correlations have been used to evaluate the life supporting capabilities of different types of stars \citep{Cuntz2016,Teal2022}, but may have systematic offsets due to the assumptions made in the \lya\ reconstructions. With a better understanding of what the intrinsic \lya\ fluxes are for main sequence stars, we can assess the accuracy of these correlations and subsequent analyses conducted with these inputs.


\section{Acknowledgements}
This research was supported by an appointment to the NASA Postdoctoral Program at the NASA Goddard Space Flight Center, administered by Universities Space Research Association through a contract with NASA. The
material includes work performed as part of the CHAMPs (Consortium on Habitability and Atmospheres of M-dwarf Planets) team, supported by the National Aeronautics and Space Administration (NASA) under grant No. 80NSSC21K0905 issued through the Interdisciplinary Consortia for Astrobiology Research (ICAR) program. ML and EWS acknowledge additional support from the Alternative Earths Team, supported by NASA under grant No. 80NSSC21K0594 issued through the ICAR program. This work made use of tools developed by the Virtual Planetary Laboratory, which is a member of the NASA Nexus for Exoplanet System Science and funded via NASA Astrobiology Program Grant No. 80NSSC18K0829.

\newpage
\appendix

\begin{figure*}[tbh]
    \centering
    \includegraphics[width=0.90\textwidth]{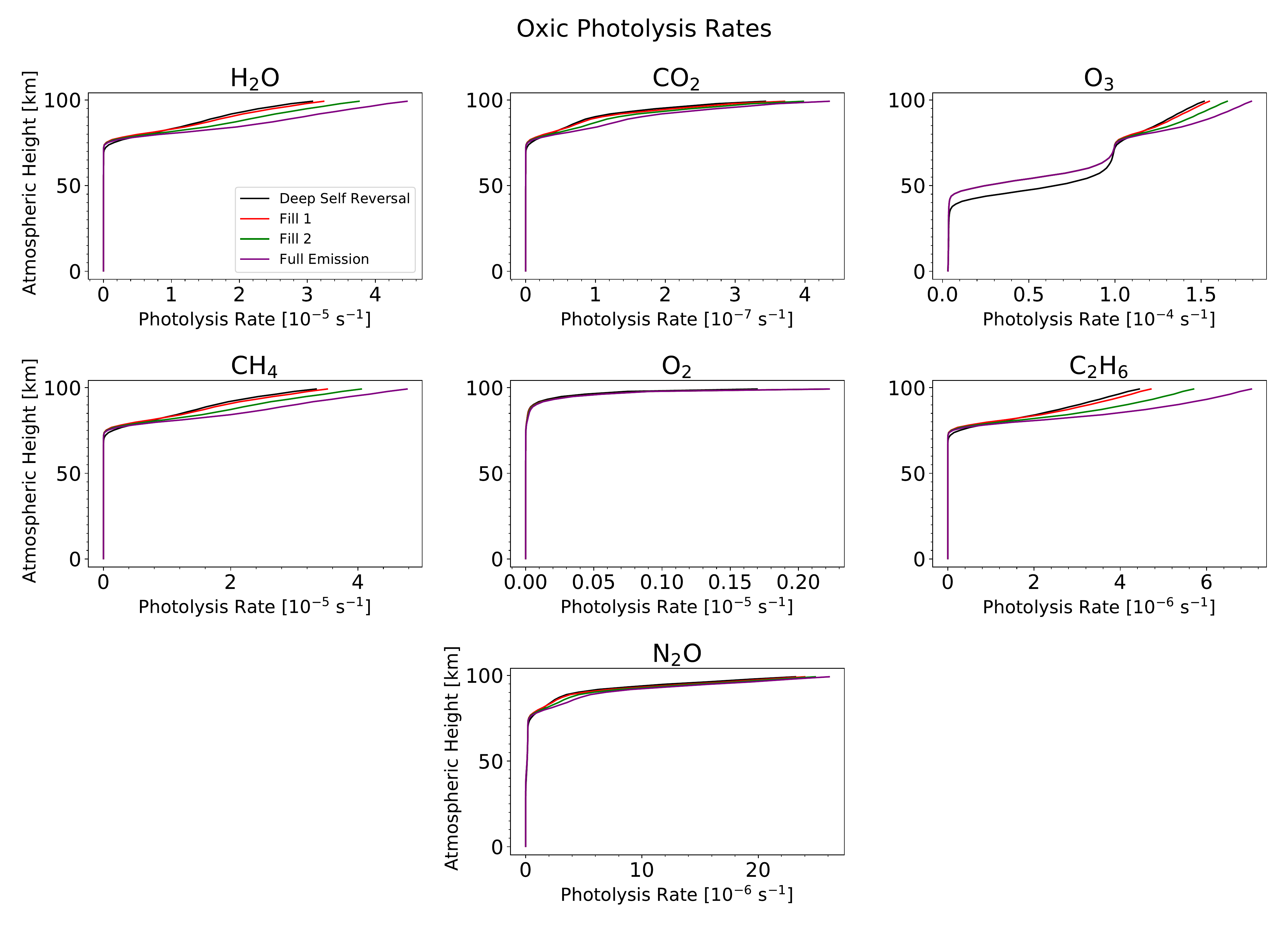}

    \caption{Comparison of photolysis rates in an oxic atmosphere for the \lya\ fill cases considered in this work. In general, photolysis rates are maximized at the top of the atmosphere and are greatest for the full emission scenario, followed by the fill 1, fill 2, and deep self-reversal cases. A linear x-axis is chosen to best illustrate the differences between the various \lya\ fill cases at the top of the atmosphere. These results correspond to the atmospheric mixing ratios shown in Figure \ref{fig:oxic_atmosphere}.
    }
    \label{fig:oxic_photolysis_rates}
\end{figure*}

\begin{figure*}[tbh]
    \centering
    \includegraphics[width=0.90\textwidth]{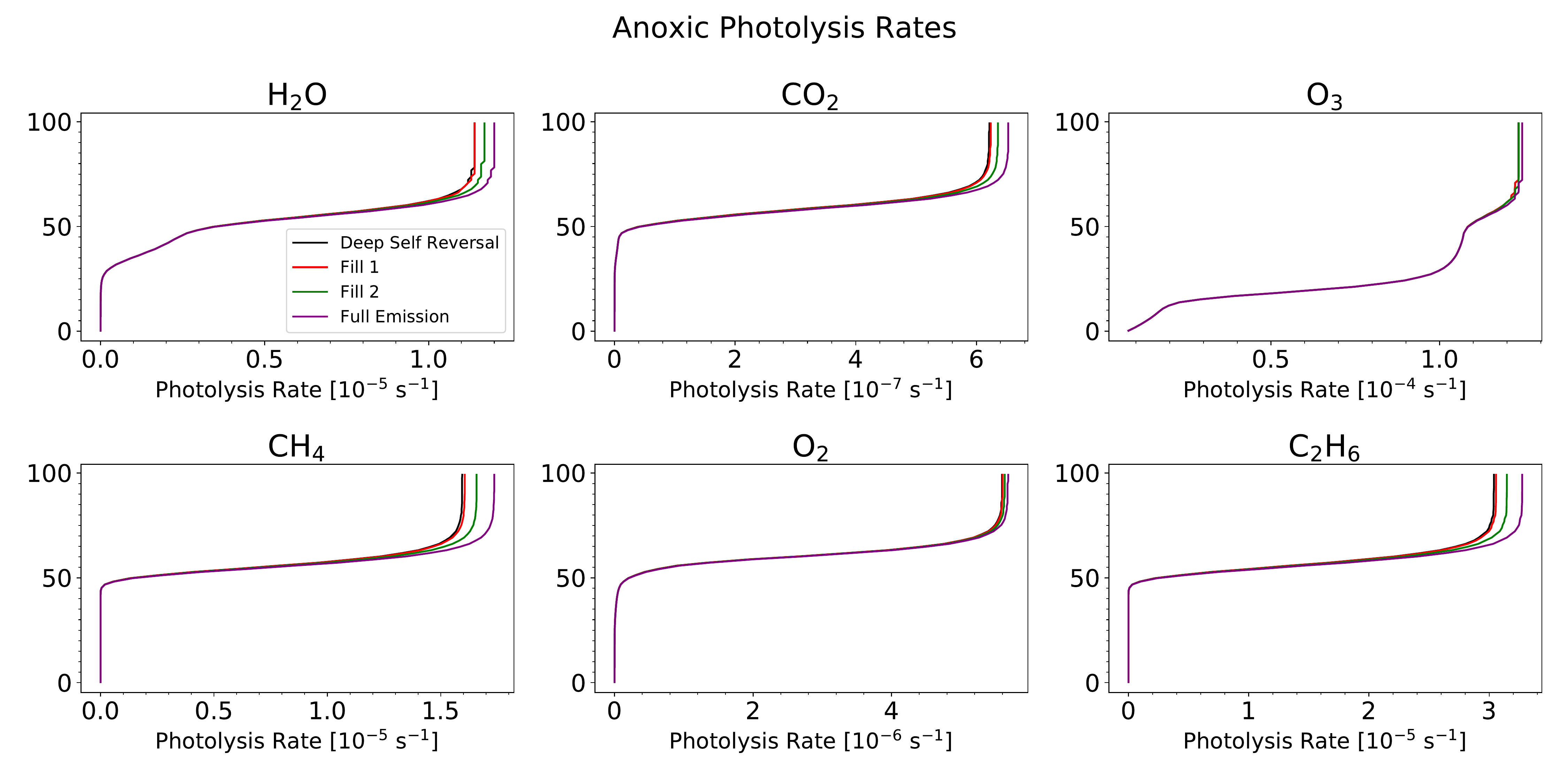}

    \caption{Same as Figure \ref{fig:oxic_photolysis_rates} but for our anoxic atmosphere scenario. These results correspond to the atmospheric mixing ratios shown in Figure \ref{fig:anoxic_atmosphere}.
    }
    \label{fig:anoxic_photolysis_rates}
\end{figure*}

As a supplement to Figures \ref{fig:oxic_atmosphere} and \ref{fig:anoxic_atmosphere} illustrating the photochemical impact of differing stellar \lya\ fluxes on oxic and anoxic terrestrial exoplanet atmospheres, we provide the corresponding altitude-dependent photolysis rates in Figures \ref{fig:oxic_photolysis_rates} and \ref{fig:anoxic_photolysis_rates}. Photolysis rates are maximized at the top of the atmosphere (mesosphere) and are greatest for the full emission case (Figure \ref{fig:bicut}, purple profile), followed by the fill 2 (Figure \ref{fig:bicut}, green profile), fill 1 (Figure \ref{fig:bicut}, red profile), and deep self-reversal (Figure \ref{fig:bicut}, black profile) cases. The photolysis rates are essentially identical in the stratosphere and troposphere for all species in both atmospheric scenarios. The only exception is the O$_{3}$ photolysis rate for the deep self-reversal case in the oxic atmosphere, though this has a minimal effect on the O$_{3}$ profile as shown in Figure \ref{fig:oxic_atmosphere}.

\newpage
\bibliography{bibliography}

\end{document}